\pdfoutput=1

\documentclass[11pt]{article}

\usepackage[final]{acl}

\usepackage{times}
\usepackage{latexsym}

\usepackage{lipsum}

\usepackage[T1]{fontenc}

\usepackage[utf8]{inputenc}

\usepackage{microtype}

\usepackage{kotex}
\usepackage[pdftex]{graphicx}
\usepackage[cmex10]{amsmath}
\usepackage{amssymb}
\usepackage{url}
\usepackage{xcolor}
\usepackage{booktabs} 
\usepackage{graphics} 
\usepackage{tabularx}
\usepackage{makecell}
\usepackage{caption}
\usepackage{subcaption}
\usepackage{balance}
\usepackage{placeins}
\usepackage{appendix}
\usepackage{multirow}
\usepackage{multicol}
\usepackage{array}
\usepackage{tabularray}
\UseTblrLibrary{booktabs}
\usepackage{enumitem}
\usepackage{adjustbox}
\usepackage{supertabular}
\usepackage{longtable}
\usepackage{multicol}
\usepackage{soul}
\newcommand{\model}{CyBERTuned}
\newcommand{\signeg}{\textcolor{red}{\textsuperscript{\textdagger}}}
\newcommand{\sigpos}{{\textsuperscript{\textdagger}}}

\newcommand{\ra}[1]{\renewcommand{\arraystretch}{#1}}

\title{Ignore Me But Don’t Replace Me: Utilizing Non-Linguistic Elements for Pretraining on the Cybersecurity Domain}

\author{
    \textbf{Eugene Jang\textsuperscript{1}} \hspace{1em}
    \textbf{Jian Cui\textsuperscript{2}\thanks{\hspace{0.15cm}Work performed while at S2W Inc.}} \hspace{1em}
    \textbf{Dayeon Yim\textsuperscript{1}} \hspace{1em} \\
    \textbf{Youngjin Jin\textsuperscript{3}} \hspace{1em} 
    \textbf{Jin-Woo Chung\textsuperscript{1}} \hspace{1em} 
    \textbf{Seungwon Shin\textsuperscript{3}} \hspace{1em}
    \textbf{Yongjae Lee\textsuperscript{1}} \vspace{0.1cm} \\
    \textsuperscript{1}S2W Inc. \hspace{1em}
    \textsuperscript{2}Indiana University Bloomington \hspace{1em}
    \textsuperscript{3}KAIST \vspace{0.1cm} \\
    \textsuperscript{1}\texttt{\{genesith,dayeon,jwchung,lee\}@s2w.inc} \\
    \textsuperscript{2}\texttt{cuijian@iu.edu} \hspace{1em}
    \textsuperscript{3}\texttt{\{ijinjin,claude\}@kaist.ac.kr}
}

\begin{document}
\maketitle
\begin{abstract}
Cybersecurity information is often technically complex and relayed through unstructured text, making automation of cyber threat intelligence highly challenging.
For such text domains that involve high levels of expertise, pretraining on in-domain corpora has been a popular method for language models to obtain domain expertise.
However, cybersecurity texts often contain non-linguistic elements (such as URLs and hash values) that could be unsuitable with the established pretraining methodologies.
Previous work in other domains have removed or filtered such text as noise, but the effectiveness of this approach has not been investigated, especially in the cybersecurity domain.
We experiment with different pretraining methodologies to account for non-linguistic elements (NLEs) and evaluate their effectiveness through downstream tasks and probing tasks.
Our proposed strategy, a combination of selective MLM and jointly training NLE token classification, outperforms the commonly taken approach of replacing NLEs.
We use our domain-customized methodology to train \model{}, a cybersecurity domain language model that outperforms other cybersecurity PLMs on most tasks.
\end{abstract}

\section{Introduction}
Cybersecurity is a critical concern as the world continues to grow reliant on technology.
Modern cybersecurity practice emphasizes the need for preemptive defense utilizing Cyber Threat Intelligence (CTI) \textemdash{} actionable information on possible cyber-threats~\cite{farnham2013tools}.
However, due to the unstructured and complex nature of such information, leveraging CTI requires extensive manual inspection by human experts~\cite{10.1145/ttpdrill}.
Although automating cyber threat intelligence has been regarded as important~\cite{cti_importance, cti_importance2}, it has been considered highly challenging~\cite{WAGNER2019101589}.

\begin{figure}[t!]
    \centering
    \includegraphics[width=.95\columnwidth]{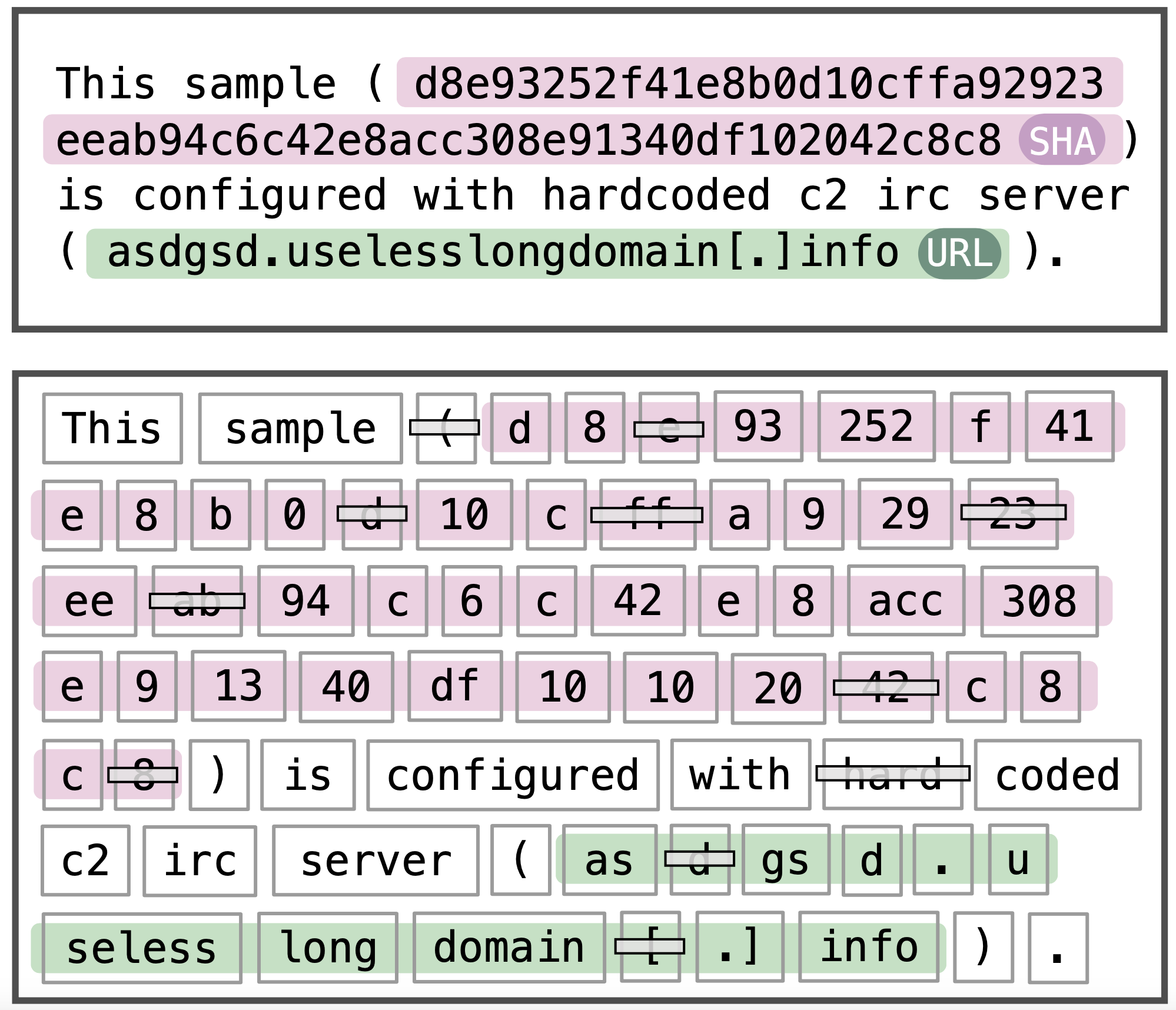}
    \caption{A threat report excerpt after tokenization and masking with 15\% probability. The tokens inside the SHA hash and URL are highlighted. Masked tokens are indicated by a gray bar.}
    \label{fig:cybersecurity-language-examples}
\end{figure}
Meanwhile, pretrained language models (PLMs) have shown great potential for text comprehension~\cite{deberta}.
However, PLMs are unlikely to have developed the necessary expertise for domains that require significant domain knowledge, such as the cybersecurity domain.
This could be somewhat addressed by extremely large models~\cite{chatgpt_ioc}, but this option is costly to train and run.
A more common approach to teach domain expertise to PLMs has been to pretrain on a domain-specific corpus.
The effectiveness of such domain-pretrained PLMs has been demonstrated in the biomedical~\cite{biobert-paper}, scientific~\cite{beltagy-etal-2019-scibert}, and legal~\cite{chalkidis-etal-2020-legalbert} domains to name a few.
Several cybersecurity domain PLMs~\cite{cybert-paper,securebert-paper, bayer2022cysecbert} were also trained in a similar manner.

However, cybersecurity texts often incorporate non-linguistic elements (NLEs) that could be inappropriate for self-supervised pretraining.
Self-supervised objectives like masked language modeling (MLM)~\cite{devlin-etal-2019-bert} or de-noising objectives~\cite{lewis-etal-2020-bart} learn by recovering original texts from a masked or modified state.
While this is mostly beneficial for natural language, such tasks might not be effective when trying to recover tokens in non-linguistic parts of text.
Figure~\ref{fig:cybersecurity-language-examples} shows an excerpt from a malware threat report containing a SHA hash and a URL.
The SHA hash tokens are linguistically random, and therefore training a model to correctly recover these tokens may not be beneficial.
Similarly, the URL tokens are less predictable compared to the natural language text surrounding it, and therefore potentially unsuitable for pretraining.

Outside of the cybersecurity domain, previous works addressed such NLEs through replacement (e.g. replace all URLs with ``\textit{[URL]}'')~\cite{social-media-bert,caselli-etal-2021-hatebert, jin-etal-2023-darkbert} or filtering~\cite{le-etal-2020-flaubert-unsupervised, t5, hung-etal-2022-ds}.
However, no attempt has been made to verify whether these approaches actually benefit pretraining.
It is also unclear whether such practices would have similar benefits in the cybersecurity domain, where it is more common for NLEs to be used alongside natural language.
Conversely, pretraining with NLEs could be beneficial to utilize the informational value of NLEs.
For instance, a model may learn to identify suspicious domains in URLs or recognize familiar hash values in the way human cybersecurity experts can.

We investigate different strategies of pretraining on the cybersecurity domain.
We first identify commonly occurring NLE types that can be extracted using regular expressions.
We then pretrain models using different MLM strategies, testing the effectiveness of selective masking and NLE token classification and comparing to the vanilla MLM and replacement strategy.
Our experiments suggest that replacement benefits on downstream tasks but harms performance on probing tasks, especially near NLEs.
Instead, we find that a strategy of selective masking while jointly training with NLE token classification generally outperforms the replacement strategy.
Using this strategy, we train \model{} (Cybersecurity BERT-like Utilizing Non-linguistic Elements of the Domain), a cybersecurity domain PLM trained with the domain-customized pretraining methodology.
We show \model{} outperforms comparable cybersecurity domain PLMs in most tasks.
\model{} model weights, training resources, and code are publicly available at \url{https://github.com/genesith/CyBERTuned}.

Our contributions are as follows:
\begin{itemize}
    \item We propose and test multiple strategies to deal with NLEs when pretraining on a cybersecurity corpus.
    \item Through experiments on a variety of domain tasks, we find a strategy that is preferable to the common practice of replacing NLEs.
    \item We use our methodology to train \model{}, a cybersecurity domain encoder model that outperforms other cybersecurity models.
    \item We provide our model weights, training resources, and code.
\end{itemize}
\section{Related Work}
\vspace{0.1cm}\noindent\textbf{Cybersecurity NLP}
Automating cyber threat intelligence has been often discussed in literature~\cite{cti_importance2, WAGNER2019101589,gapfinder}.
Classical off-the-shelf NLP methods, such as regex processing and dependency parsing, have been used to extract attack patterns~\cite{10.1145/ttpdrill} or malware behaviors~\cite{10.1145/featuresmith} from cybersecurity texts.
Recent works explore the potential of using BERT~\cite{devlin-etal-2019-bert} for more complex tasks such as exploitability prediction~\cite{YIN2020106529}, malware detection~\cite{rahali2021malbert}, and dark web analysis~\cite{jin-etal-2023-darkbert}.

\begin{figure*}[th!]
    \centering
    \includegraphics[width=\linewidth]{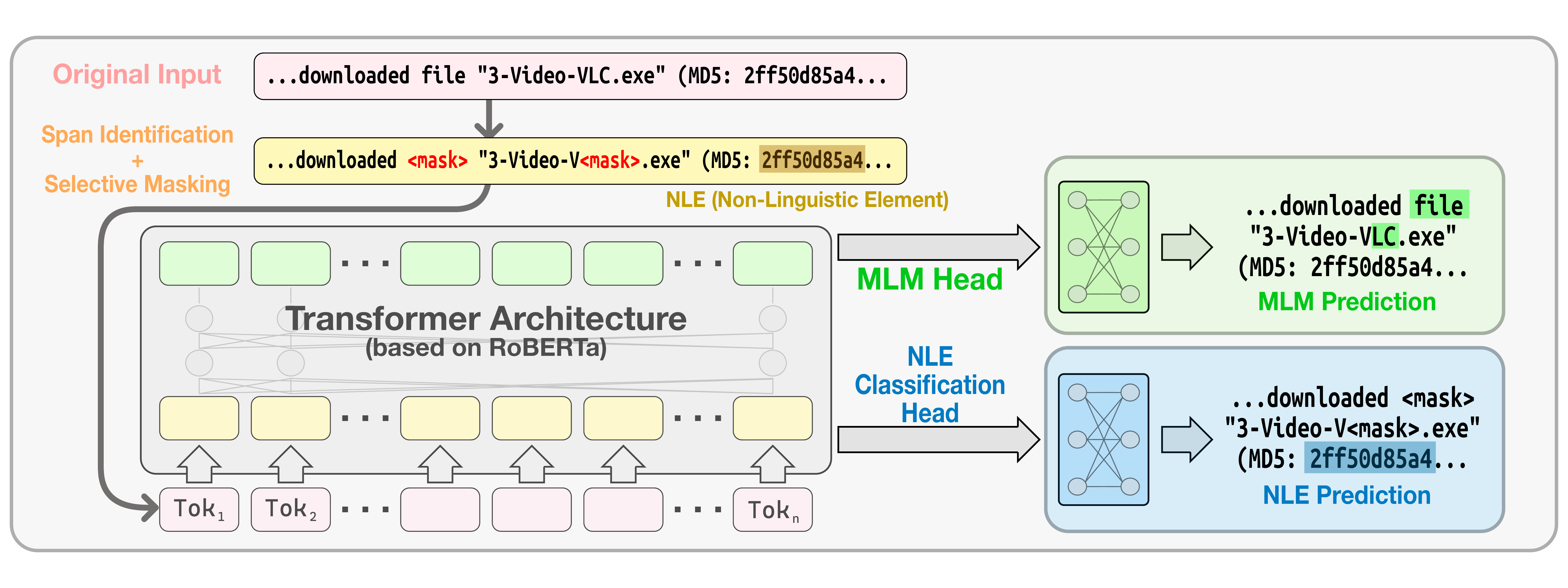}
    \caption{The overall architecture of \model{}. NLE spans from the original input is used in the masking step and for NLE classification.}
    \label{fig:Architecture}
\end{figure*}
\vspace{0.1cm}\noindent\textbf{Domain PLMs}
PLMs train with large text corpora using self-supervision tasks~\cite{devlin-etal-2019-bert, lewis-etal-2020-bart, deberta}.
Many domain PLMs~\cite{biobert-paper, chalkidis-etal-2020-legalbert, beltagy-etal-2019-scibert} were able to outperform general PLMs on domain-specific tasks by simply replicating existing pretraining procedures on domain corpora.
PLMs for the cybersecurity domain using this approach have been suggested by several works~\cite{cybert-paper, securebert-paper, bayer2022cysecbert}. Our work differs in that we use a domain-customized methodology after investigating the effectiveness of various strategies.

\vspace{0.1cm}\noindent\textbf{Pretraining Strategies}
Self-supervised tasks for pretraining have been investigated by many works~\cite{lewis-etal-2020-bart,aroca-ouellette-rudzicz-2020-losses,yamaguchi-etal-2021-frustratingly}.
Some works find improvements by modifying the masking procedure of the MLM task.
Changing masking from token-level to word-level significantly improved BERT pretraining for Chinese~\cite{Chinese-wwm}.
Works on selective masking suggested that masking tokens important to tasks~\cite{gu-etal-2020-train-no-evil}, entities~\cite{lin-etal-2021-entitybert}, or reasoning~\cite{sanyal-etal-2023-apollo} more frequently was effective.
Conversely, our selective masking method skips masking tokens that are ineffective for MLM training.
Previously, dropping less important tokens mid-training was suggested as a way to increase training efficiency~\cite{hou-etal-2022-token}, but at the cost of losing semantic sensitivity~\cite{zhong-etal-2023-revisiting}.
In our work, ineffective tokens are identified beforehand via NLE spans, and are skipped during masking rather than dropped during training.

\section{Method}
In this section, we first discuss the types of NLEs in cybersecurity texts and how to extract NLE instances.
We then propose some methods to utilize the extracted NLE spans in pretraining with cybersecurity texts.


\subsection{Non-linguistic Elements}
\label{subsec:method-nles}
Cybersecurity texts often feature non-linguistic text alongside natural language. 
Among non-linguistic texts, certain types of are extractable with regular expressions~\cite{10.1145/ttpdrill}.
We narrow our scope to non-linguistic elements that can be identified by regular expressions, since our aim is to apply them into self-supervised tasks.
After manual inspection of cyber threat reports, we select the NLE types that are both frequent and identifiable with regular expressions.
The following 7 types were selected: \textit{URLs}, \textit{email addresses}, \textit{IP addresses}, \textit{MD5 hashes}, \textit{SHA hashes}, \textit{Bitcoin addresses}, and \textit{CVE IDs}\footnote{CVE (Common Vulnerabilities and Exposures) IDs are unique identifiers assigned to publicly disclosed vulnerabilities.}.
Note that we do not consider NLE types that require significant effort to extract precisely, such as filepaths or code blocks.
We also extend detection to defanged NLEs (e.g., \textit{hxxp}://example.com, 192.168\textit{[.]}1.192) by utilizing the iocide Python library\footnote{\url{https://pypi.org/project/iocide/}}.

\subsection{Leveraging NLE Spans in Pretraining}
\begin{table*}[t]
    \renewcommand{\multirowsetup}{\centering}
    \centering
    \definecolor{beaublue}{rgb}{0.77, 0.87, 0.9}
    \definecolor{grn}{rgb}{0.0, 0.0, 0.8}
    \definecolor{lvd}{rgb}{0.89, 0.02, 0.17}
    \sethlcolor{beaublue}

    \resizebox{\textwidth}{!}{ 
        \begin{tabular}{lccc>{\raggedright\arraybackslash}p{0.38\textwidth}  >{\raggedright\arraybackslash}p{0.38\textwidth}}
        \toprule
        \multicolumn{1}{c}{\multirow{2}{*}{Strategies}} & \multicolumn{2}{c}{Masks} & \multicolumn{1}{c}{\multirow{2}{*}{NLEC}} & \multicolumn{1}{c}{\multirow{2}{*}{\makecell{Example\\(MLM)}}} & \multicolumn{1}{c}{\multirow{2}{*}{\makecell{Example\\(NLE Classification)}}} \\
        \cmidrule{2-3}
         & SLE & FNLE \\ 
        \midrule
        \multirow{1}{*}{Vanilla MLM}   & \multirow{1}{*}{\checkmark} &  \multirow{1}{*}{\checkmark} &  & \makecell[l]{ \hl{The Dropper drops a zipped SysJoker} \\ \hl{(53f1bb23f670d331c9041748e7e8e396}\\ \hl{) from C2 https{[}://{]}github{[}.{]}url-mini{[}.{]}}\\\hl{com/msg.zip, copies it to} } & \makecell[c]{N/A}
        \\
        \midrule
        \multirow{1}{*}{Replace All} & \multicolumn{3}{c}{\multirow{1}{*}{(replaced)}}  & \makecell[l]{ \hl{The Dropper drops a zipped SysJoker} \\ \hl{(\textless{}MD5\textgreater{}) from C2 \textless{}URL\textgreater{}, copies it to} } & \makecell[c]{N/A}\\
        \midrule
        \multirow{1}{*}{Vanilla + NLEC}   & \multirow{1}{*}{\checkmark} & \multirow{1}{*}{\checkmark}  & \multirow{1}{*}{\checkmark} & \makecell[l]{ \hl{The Dropper drops a zipped SysJoker} \\ \hl{(53f1bb23f670d331c9041748e7e8e396}\\ \hl{) from C2 https{[}://{]}github{[}.{]}url-mini{[}.{]}}\\\hl{com/msg.zip, copies it to} } & \makecell[l]{ The Dropper drops a zipped SysJoker \\ ({\color{lvd}53f1bb23f670d331c9041748e7e8e396} \\ ) from C2 {\color{grn}https{[}://{]}github{[}.{]}url-mini{[}.{]} } \\ {\color{grn}com/msg.zip}, copies it to}
          \\
        \midrule
        \multirow{1}{*}{Mask-Semis} & \multirow{1}{*}{\checkmark} &     &  & \makecell[l]{ \hl{The Dropper drops a zipped SysJoker} \\ \hl{(}53f1bb23f670d331c9041748e7e8e396 \\ \hl{) from C2 https{[}://{]}github{[}.{]}url-mini{[}.{]}} \\ \hl{com/msg.zip, copies it to} } & \makecell[c]{N/A}
        \\
        \midrule
        \multirow{1}{*}{Mask-Semis + NLEC} & \multirow{1}{*}{\checkmark} &     & \multirow{1}{*}{\checkmark} & \makecell[l]{ \hl{The Dropper drops a zipped SysJoker} \\ \hl{(}53f1bb23f670d331c9041748e7e8e396 \\ \hl{) from C2 https{[}://{]}github{[}.{]}url-mini{[}.{]}} \\ \hl{com/msg.zip, copies it to} } & \makecell[l]{ The Dropper drops a zipped SysJoker \\ ({\color{lvd}53f1bb23f670d331c9041748e7e8e396} \\) from C2 {\color{grn}https{[}://{]}github{[}.{]}url-mini{[}.{]} } \\ {\color{grn}com/msg.zip}, copies it to} \\
        \midrule
        \multirow{1}{*}{Mask-None + NLEC} &  &     & \multirow{1}{*}{\checkmark} &  \makecell[l]{ \hl{The Dropper drops a zipped SysJoker} \\ \hl{(}53f1bb23f670d331c9041748e7e8e396\\ \hl{) from C2} https{[}://{]}github{[}.{]}url-mini{[}.{]}\\com/msg.zip\hl{, copies it to} } & \makecell[l]{ The Dropper drops a zipped SysJoker \\ ({\color{lvd}53f1bb23f670d331c9041748e7e8e396} \\) from C2 {\color{grn}https{[}://{]}github{[}.{]}url-mini{[}.{]} } \\ {\color{grn}com/msg.zip}, copies it to} \\

        \bottomrule
        \end{tabular}
    }
    \caption{Comparisons between how each text types are processed in different strategies. In the MLM examples, highlighted sections indicate text that are considered for masking. In the NLE Classification examples, each token is predicted for its NLE type (indicated by color).}
    \label{tab:PretrainExample}
    \bigskip\bigskip

\end{table*}

We study two methods to leverage extracted NLE spans to guide the pretraining. Figure~\ref{fig:Architecture} shows a model utilizing both methods.

\vspace{0.1cm}\noindent\textbf{NLE Classification}: The model is explicitly instructed to predict which tokens belong to NLEs in the pretraining text.
This can be modeled as a simple token classification task and can be trained alongside the MLM task.
Each token is labeled with its NLE type (0 if outside of NLE span), which is predicted by a token classification head (linear layer).

Since this task is a more semantically shallow task compared to the original MLM task, it should not dominate the total loss function~\cite{aroca-ouellette-rudzicz-2020-losses}.
Therefore, we apply a scaling factor (0.1) before adding with the MLM loss to produce the total loss.

\vspace{0.1cm}\noindent\textbf{Selective MLM}: In the masking stage, the NLE spans are used to avoid masking tokens that are inside NLEs.
However, since informational content varies between NLE types it must be investigated whether all NLE types should be avoided.
We note that NLEs that involve human generated text (URLs and emails), unlike protocol-generated values (IP addresses, hash values, etc.), can contain linguistically meaningful information. 
For instance, a human expert may identify the URL \textit{github[.]url-mini[.]com/msg.zip} as a malicious file download link from a fake domain masquerading as the legitimate GitHub domain.

To make a simple distinction between NLE types, we group the NLEs based on whether they are generated by humans or protocol.
Specifically, we group \textit{URLs} and \textit{emails} as \textbf{semi-linguistic elements} (SLEs) and \textit{IP addresses}, \textit{MD5 hashes}, \textit{SHA hashes}, \textit{Bitcoin addresses}, and \textit{CVE IDs} as \textbf{fully non-linguistic elements} (FNLEs).
We then test two settings of selective masking:
\textbf{Mask-None}: all NLE types are avoided during masking.
\textbf{Mask-Semis}: fully non-linguistic elements are avoided but semi-linguistic NLEs are allowed to be masked.
\section{Pretraining the Models}
\label{sec:pretrain}
To evaluate our pretraining methodology on the cybersecurity domain, we pretrain models on a cybersecurity text corpus using a number of strategies.
We first describe the tested pretraining strategies, including the vanilla MLM and ablation settings.
We also describe our cybersecurity corpus and show statistics that suggest NLEs are more frequent in cybersecurity texts.
\subsection{Pretraining Strategies}

We compare a total of 6 pretraining strategies. First we include two commonly used strategies as baselines. As ablation studies, we also test two strategies using only one method. Then two strategies that utilize both NLE classification and selective MLM are described.
Examples of these strategies can be seen in Table~\ref{tab:PretrainExample}.

\vspace{0.1cm}\noindent\textbf{Vanilla MLM:} The original masking strategy~\cite{devlin-etal-2019-bert}. After tokenization, 15\% of the input tokens are selected for prediction. Following the original implementation, 80\% are converted into the mask token, 10\% are converted into a random token, and 10\% are unchanged.

\vspace{0.1cm}\noindent\textbf{Replace All:} A commonly used strategy to reduce the impact of NLEs in pretraining~\cite{caselli-etal-2021-hatebert, jin-etal-2023-darkbert}. The MLM method is unchanged, but the NLEs in the input corpus is converted to an identifier of the NLE type (e.g., all CVE IDs are replaced with ``<CVE>''). However, comes with a risk of reducing informational content in the pretraining corpus.

\vspace{0.1cm}\noindent\textbf{Vanilla + NLEC:} The MLM method is unchanged, but the joint task of token-level NLE classification (NLEC) is also performed. While MLM is still done on NLE tokens, NLEC could instruct the model to understand the different role the tokens have.

\vspace{0.1cm}\noindent\textbf{Mask-Semis:} A selective MLM method that avoids masking of FNLEs (hash values, IP addresses, etc.) while allowing masking of SLEs (URLs and emails).

\vspace{0.1cm}\noindent\textbf{Mask-Semis + NLEC:} A strategy using both the selective masking and NLEC. In this setting, tokens in SNLEs are allowed to be masked and tokens in FNLEs are avoided.

\vspace{0.1cm}\noindent\textbf{Mask-None + NLEC:} A strategy using both the selective masking and NLEC. In this setting, tokens in all NLEs are avoided during masking.

\subsection{Cybersecurity Corpus}
\begin{table}[t]
    \centering
    \scalebox{0.9}{
      \begin{tabular}{lrr}
        \toprule
        \textbf{Data Source} & \textbf{Count} & \textbf{Data Size} \\
        \midrule
        \textit{Full}\\
        Online security articles & 150K & 680.8 MB\\
        Security paper abstracts & 7.3K & 9.1 MB\\ 
        Wikipedia articles & 3.4K & 15.7 MB\\ 
        CVE descriptions & 185K & 52.5 MB\\ 
        \midrule
        \textit{Pretraining Subset}\\
        Online security articles & 34K & 170.4 MB\\
      \bottomrule
    \end{tabular}%
    }
    \caption{Statistics of data sources used in the corpus. The data used to pretrain the models is a subset of the total data.}
    \label{tab:datacollection-statistics}
\end{table}
We collect and curate a large amount of text from publicly available online sources. Like other cybersecurity PLMs, we construct our corpus with data from a variety of sources: Online Security Articles, Security Paper Abstracts, Wikipedia Articles, and CVE Descriptions.
A detailed description of the components and collection of the corpus can be found in Appendix~\ref{appendix:corpus} and~\ref{sec:articles-sources}.

\vspace{0.1cm}\noindent\textbf{Pretraining subset.} We further identify a subset of the corpus focused on threat reports to pretrain on.
This is because the full corpus covers a variety of styles, including news articles written for non-expert audiences.
Such articles contain little technical information and few NLEs.
Since our goal is to compare pretraining strategies of teaching technical expertise of analysts to models, we filter to find sources that publish for expert audiences.
From 60 total online source sites, we select 30 sites that more often feature technical information to make up the pretraining subset.
The data size of our sources used for constructing the corpus and the pretraining subset is listed in Table~\ref{tab:datacollection-statistics}.

\vspace{0.1cm}\noindent\textbf{NLE Statistics.}
To demonstrate the frequency of non-linguistic elements in our cybersecurity text corpus, we compare our corpus with two general domain text corpora: the Wikipedia corpus and the C4 corpus~\cite{t5}.
The Wikipedia corpus, used in pretraining BERT and other models, consists of text content from Wikipedia articles.
The C4 corpus, first used for pretraining T5~\cite{t5}, is a collection of crawled web pages. Unlike our corpus, the C4 corpus aims to include only natural language text and use heuristics to filter text with non-natural language.
Due to the large size of this dataset, we sample 0.1\% of the total size (365,234 documents) for our analysis.

To compare between corpora, we calculate the frequency of NLEs.
We first count the number of instances of each NLE type, using our detection methodology (discussed in Section~\ref{subsec:method-nles}) on each corpus.
We use the NLTK~\cite{bird-loper-2004-nltk} tokenizer to count the number of words in each corpus.
Table~\ref{tab:statistics} shows the frequencies of each non-linguistic element per million words.
We observe that the frequency of NLEs in our corpus is significantly higher compared to the two general domain corpora.

\begin{table}[t]
    \centering
    \scalebox{0.9}{
      \begin{tabular}{l|rrrr}
        \toprule
        \textbf{NLE} & \makecell{\textbf{Ours}\\ \textbf{(PS)}}&\makecell{\textbf{Ours}\\ \textbf{(Full)}} &  \textbf{Wiki} & \textbf{C4} \\
        \midrule
        URL & 16,272 &5,172 & 62 & 404 \\
        EMAIL & 3,282 &901& < 1    & 33 \\
        IP & 2,503    &780& 3  & 15 \\
        MD5 & 2,651   &754& < 1     & 1 \\
        SHA & 550   &161& < 1     & < 1 \\
        BTC & 1,024   &273& < 1      & < 1 \\
        CVE & 1,225   & 550&< 1     & 3 \\
        \bottomrule
      \end{tabular}%
      }
      \caption{Distribution of non-linguistic elements (per million words) in our pretraining subset (PS) corpus, full corpus, Wikipedia, and C4.}
      \label{tab:statistics}
\end{table}
\begin{table*}[th!]
\small
\centering
\ra{1.3}
\resizebox{0.85\linewidth}{!}{
    \begin{tabular}{lcccccc}
    \toprule
    \multirow{2}{*}{Strategies} & \multicolumn{3}{c}{Downstream Tasks} && \multicolumn{2}{c}{Probing Tasks} \\
    \cmidrule{2-4} \cmidrule{6-7} 
    & CyNER & CySecED & MTDB && All & Near-FNLEs \\ 
    \midrule
     RoBERTa &
     0.637 & 0.504 & 0.802 &&
     0.278 & 0.270 \\
     Vanilla MLM &
     0.648 & 0.510 & 0.822\sigpos &&
     0.382 & 0.460 \\
     Replace All &
     \underline{0.664\sigpos*} & \underline{0.544} & \underline{0.827\sigpos} &&
     0.381 & 0.438 \\
     Vanilla + NLEC &
     0.652 & 0.526 & 0.820 &&
     0.380 & 0.455 \\
     Mask-Semis &
     0.638 & 0.538 & 0.817 &&
     \textbf{0.386} & \underline{0.463} \\
     Mask-Semis + NLEC &
     \textbf{0.667\sigpos*} & \textbf{0.544\sigpos} & 0.825\sigpos && 
     \ul{0.383} & \textbf{0.464}\\
     Mask-None + NLEC &
     0.643 & 0.533\sigpos & \textbf{0.829\sigpos} &&
     0.382 & 0.452 \\
    \bottomrule
    \end{tabular}
}
\caption{Experimental results on multiple pretraining strategies. Downstream tasks show median values over 10 runs. \textbf{Boldface} represents the best score and \underline{underlined} values represents the second best score.
The \textsuperscript{\textdagger} symbols indicates statistically significant distributions from the RoBERTa-base baseline.
The * symbols indicates statistically significant distributions from the Vanilla MLM baseline.}
\label{tab:PretrainRes}
\end{table*}
\subsection{Pretraining Setup}
For our experiments, we pretrain further on the pretrained RoBERTa-base model~\cite{roberta-paper}.
We choose the RoBERTa model as the base architecture because its minimal pre-tokenization scheme and coverage is suitable to our corpus\footnote{The BERT pretokenizer assumes there are spaces between the `:', `/', `.' characters common in URLs. The corpus also contains obscure characters that aren't considered by other tokenizers(the T5 tokenizer does not have the `\symbol{92}' character in its vocabulary.}.
For efficiency, we choose to pretrain further on the pretrained model, following findings that suggest that this method is as effective as training a model from scratch~\cite{chalkidis-etal-2020-legalbert, el-boukkouri-etal-2022-train-scratch}. We mostly follow RoBERTa's training hyperparameters, with few modifications to account for our smaller corpus size (details can be found in Appendix~\ref{appendix:Experiment-settings}).
Note that the \textit{Replace All} strategy modifies the pretraining corpus size.
For fair comparison, all models were trained for 500 steps ($\sim$12 epochs for the \textit{Replace All }model, $\sim$10 epochs for other models).
\section{Experiments}
\label{sec:experiments}

We evaluate the models trained by each pretrained strategy with both downstream tasks and probing tasks. For comparison, we also experiment with the base RoBERTa model.
\subsection{Downstream Tasks}
We compare the ability of each model to fine-tune onto downstream tasks using challenging cybersecurity datasets.

\vspace{0.1cm}
\noindent\textbf{CyNER}~\cite{cyner}: A named entity recognition dataset of annotated malware threat reports. The reports are annotated for five entity types: Malware, System, Organization, Indicator, and Vulnerability. 

\vspace{0.1cm}
\noindent\textbf{CySecED}~\cite{cyseced-paper}:
An event detection dataset of annotated articles from The Hacker News.
The articles are annotated for 30 fine-grained events types describing cyber-attacks or vulnerabilities.

\vspace{0.1cm}
\noindent\textbf{MalwareTextDB (MTDB)}~\cite{lim-etal-2017-malwaretextdb, phandi-etal-2018-semeval}:
A dataset of malware reports annotated for four types of attributes ActionName, Capability, StrategicObjectives and TacticalObjectives.
The labels are cast into a multiple choice question format, where the objective is to identify the correct attribute given a passage, attribute type, and answer choices.

\subsection{Probing Tasks}
A disadvantage of comparing performance with downstream tasks is that fine-tuning modifies the model weights learned from pretraining.
In order to evaluate the model weights themselves, we probe the model's ability to produce correct MLM answers for relevant tokens similar to the LAMA~\cite{petroni-etal-2019-lama} framework.
We follow the domain-specific version by \citet{chalkidis-etal-2023-legallama}, in which a list of legal terminology was used to find instances of the terms from a target corpus.
Models are then evaluated by its ability to recover the correct terminology after it is masked.

Our implementation tests model ability to correctly identify cybersecurity terminology in text context.
We first construct a list of relevant terminology by taking words used in MITRE's database of enterprise attack techniques\footnote{\url{https://attack.mitre.org/techniques/enterprise/}}.
After processing and filtering, we identify 226 tokens to be used for probing (see Appendix~\ref{appendix:probing}).
To probe the ability of each model, we evaluate the MLM performance on the validation split of the full corpus after masking all target tokens. A total of 77,983 tokens were masked. We also mark tokens in the vicinity (within 20 tokens away) of FNLEs, to see if the presence of FNLEs affects the probing performance. A total of 4,906 tokens were near FNLEs.



\subsection{Results}
The results of the experiments are presented in Table~\ref{tab:PretrainRes}. For downstream tasks, we report the median values over 10 seeds. We mark statistical significance of $p<0.05$ compared with the base RoBERTa and \textit{Vanilla MLM} baselines.
F1 scores are shown for the CyNER and CySecED tasks and accuracy is shown for the MTDB task and probing tasks.

\noindent\textbf{NLEC.} Comparing the \textit{Vanilla MLM} with the \textit{Vanilla + NLEC model} suggests that the classification task, on its own, does not provide meaningful benefits.
However, when comparing the \textit{Mask-Semis} and \textit{Mask-Semis + NLEC} settings, the addition of the NLEC task provides a noticeable benefit in downstream tasks. In both comparisons, NLEC caused a slight decrease in the probing tasks.

\noindent\textbf{Selective Masking.} Comparing the \textit{Vanilla MLM} with the \textit{Mask-Semis} model suggest that selective masking does not produce consistent gains in downstream tasks, although it benefits probing tasks.
Comparing \textit{Mask-Semis + NLEC} and \textit{Mask-None + NLEC} settings, the strategy of masking SLEs seems to benefit more consistently across downstream tasks and the probing tasks. This results suggests that there is value in performing masking on URLs and emails.

\noindent\textbf{Best performers.} While different pretraining methods suit different tasks~\cite{lewis-etal-2020-bart}, the \textit{Mask-Semis + NLEC} model performed consistently well across all tasks.
The \textit{Replace All} model was also very capable in downstreaming tasks, but was weaker in probing tasks.
Especially, the model probing performance was worst of all the pretrained models when the probed token was near an FNLE.
This is an undesirable characteristic of the model, since the model is expected to encounter multiple FNLEs in the domain.
We argue \textit{Mask-Semis + NLEC} is the best strategy because it allows the model to utilize NLEs while achieving high downstream performance.
\section{\model{} Experiments}
\subsection{Pretraining \model{}}
\label{sec:cybert}
With our findings, we train our final model \model{}.
We train on a larger scale with the \textit{Mask-Semis + NLEC} strategy. We compare our model with other language models on a larger array of downstream tasks in the cybersecurity domain.
The \model{} model is trained on our full cybersecurity corpus using a similar setup. Compared to the previous experiments, we train longer for a total of 200 epochs on a larger corpus.
\begin{table*}[h!]
    
    \small
    \centering 
   
    \ra{1.5}
    \begin{adjustbox}{width=.9\textwidth}
    \begin{tabular}{lcccccccc}
        \toprule
     \multirow{2}{*}{} & \multicolumn{3}{c}{Token Class.} &&\multicolumn{2}{c}{Sequence Class.} &&
     \multicolumn{1}{c}{MCQA}\\
    \cmidrule{2-4} 
    \cmidrule{6-7}
    \cmidrule{9-9}
        
                         & CASIE
                         & CyNER              & CySecED           &               & CYDEC  & TT             && MTDB              \\
        \midrule
        RoBERTa-base&0.748& 0.637& 0.504&&
        \ul{0.829}& 0.831&& 
        0.802\\
        CyBERT&0.711\signeg{}& 0.462\signeg{}& 0.361\signeg{}&&
        0.798& 0.832&& 
        0.731\signeg{}\\
        CySecBERT&0.734& 0.572\signeg{}& 0.491&&
        0.814& \ul{0.845}\sigpos{}&& 
        0.808\\
        SecureBERT&\textbf{0.753}& \ul{0.638}& \ul{0.529}\sigpos{}&&
        0.816& 0.828&& 
        \ul{0.825}\\
        \model{} (Ours)&\ul{0.750}& \textbf{0.654}& \textbf{0.585}\sigpos{}&&
        \textbf{0.844}& \textbf{0.857}\sigpos{}&& 
        \textbf{0.861}\sigpos{}\\
        \bottomrule
    \end{tabular}%
    \end{adjustbox}
    
    \caption{Experimental results of \model{} and baselines on downstream cybersecurity tasks, showing median F1 scores across 10 seeds. \textbf{Boldface} values represents the best score and \underline{underlined} values represents the second best score. The symbols \sigpos{} (positive) and \signeg{} (negative) indicates statistically significant distributions from the baseline (RoBERTa-base).}
    \label{tab:FinalExp}
\end{table*}

\subsection{Downstream Tasks}
\label{sec:full-datasets}

We conduct downstream tasks\footnote{Note that we do not do the probing tasks, since only our model was trained on the same sources of text with the validation corpora.} on a wider variety of cybersecurity tasks.
The new tasks are described below.

\vspace{0.1cm}
\noindent\textbf{CASIE}~\cite{satyapanich2020casie}:
An event detection dataset of annotated news articles for non-expert audiences.
The articles are annotated for five event types: data breach, phishing, ransom, discover, and patch.

\vspace{0.1cm}
\noindent\textbf{TwitterThreats (TT)}~\cite{zong-etal-2019-analyzing}:
A binary sequence classification dataset of annotated tweets that mention threat keywords.
Each tweet is annotated on whether the tweet describes a threat to the mentioned entity.

\vspace{0.1cm}
\noindent\textbf{CYDEC}~\cite{yagcioglu-etal-2019-cydec}:
A binary sequence classification dataset of annotated tweets that mention cybersecurity keywords.
Each tweet annotated on whether the tweet describes a cybersecurity-related event.


\subsection{Baselines}
We compare \model{} with the base RoBERTa model and other cybersecurity domain PLMs. All models follow the 12-layer Transformer encoder architecture.

\vspace{0.1cm}
\noindent\textbf{RoBERTa-base}~\cite{roberta-paper}:
The RoBERTa-base model that was used to initialize \model{}.

\vspace{0.1cm}
\noindent\textbf{CyBERT}~\cite{cybert-paper}:
A cybersecurity BERT-based model, further pretrained on the base BERT model. The BERT vocabulary is extended by 1,000 tokens from the training corpus identified by TF-IDF.
    
\vspace{0.1cm}
\noindent\textbf{CySecBERT}~\cite{bayer2022cysecbert}:
A cybersecurity BERT-based model, further pretrained on the base BERT model. The model uses the BERT vocabulary.

\vspace{0.1cm}
\noindent\textbf{SecureBERT}~\cite{securebert-paper}:
A cybersecurity RoBERTa-based model, further pretrained on the base RoBERTa model.
The model uses a custom vocabulary with adjusted token weights.


\subsection{Results}
The experiment results are presented in the Table~\ref{tab:FinalExp}.
As before, we conduct each experiment over 10 seed values and report median values.

The base RoBERTa model, despite being pretrained on the general domain, performed better than some domain PLMs in several tasks.
Of the two BERT-based models, CyBERT performed poorly on most tasks while CySecBERT generally showed competitive performance. However, even CySecBERT performed poorly in the CyNER task. This is possibly due to its usage of the uncased BERT tokenizer, which not only distorts texts with special characters but is case-insensitive (possibly important for NER).

SecureBERT was the only model that beat \model{} in a task. It showed high performance in the token-level tasks, suggesting some benefit of their custom tokenizer. On the other hand, \model{} performed consistently, achieving best or second-best performance in all tasks.

\section{Discussion}
\label{sec:discussion}
\noindent{\textbf{RoBERTa's performance.}}
The base RoBERTa model achieved good performance on certain tasks after fine-tuning, often outperforming domain-pretrained models.
A possible interpretation is that previously challenging cybersecurity tasks, such as binary sequence classification of threat tweets, do not require extensive domain knowledge to achieve high performance.
It should be noted that the CYDEC dataset reports a human F1-score of 0.59 and the TwitterThreats dataset reports a Cohen’s κ of 0.66, suggesting these models have possibly already exceeded human and annotator performance on these tasks.
This underscores the need for more challenging datasets for benchmarking models in the cybersecurity domain.

\vspace{0.1cm}
\noindent{\textbf{NLE Classification as auxiliary task.}} 
Although MLM loss takes long to plateau, NLE classification loss plateaus quite early during pretraining.
We note that the NLE classification overhead is not large, training with and without NLE classifications only had a 0.6\% difference in training time.
One possible method to increase training efficiency might be to drop NLE classification task after the loss plateaus. Whether this would achieve comparable performance could be investigated in further work.

\vspace{0.1cm}
\noindent{\textbf{NLEs of other domains.}} While it is common practice to remove NLEs in other domains, our investigation suggests proper modifications to training may be preferable.
However, our experiments were conducted in the cybersecurity domain, where certain NLE types can contain informative content.
The optimal pretraining strategy is likely different across text domains, and dependent on informational content of NLEs.
\section{Conclusion}
We investigate methods to modify pretraining to suit the cybersecurity domain.
We find that a strategy of selective MLM that allows for masking of semi-linguistic elements but not fully-linguistic elements with an auxiliary NLE classification task showed best performance.
With these findings, we present \model{}, a cybersecurity PLM with our modified pretraining methodology.
The final \model{} model shows strong performance across all cybersecurity downstream tasks.
Our findings support the importance of adapting pretraining methodologies to suit target domains.
\section*{Limitations}
\label{sec:limits}
\noindent{\textbf{Non-linguistic Element Types.}} The types of non-linguistic element discussed in this work represent a subset of a large set of textual data that are atypical to natural language.
While there are more text types that fall under this category, our scope was limited to types that are easily identifiable to be practical for self-supervision.
We note the existence of more complex types of text that are relevant to understanding cybersecurity text such as code blocks, filepaths, or log entries.
We leave the detection and effective utilization of such text types for future work.

Another limitation is that the strategies tested utilized broad distinctions between SLEs and FNLEs. Due to computational restraints, it was not possible to pretrain while treating each NLE type uniquely.
Therefore, even with our method that utilizes masking SLEs to improve performance, it is difficult to attribute the performance gains to specific individual NLE types.
For now we focus discussions on our empirical results, and leave fine-grained analysis to future work.

\noindent{\textbf{Downstream tasks.}} There are many factors to consider when fine-tuning downstream tasks across multiple models. We attempt to find stable settings that allow all runs to be successful, but there are inconsistent runs.
To mitigate this, we report median values of 10 seed values and suggest the probing task as an alternative.
Since our experiments are run across multiple models, multiple tasks, and multiple hyperparameters, there may be cases of novel untested hyperparameter combinations on model-task combinations that have not been explored.

\vspace{0.1cm}
\noindent{\textbf{NLEs in different domains.}} As stated in Section~\ref{sec:discussion}, the findings in this work were investigated only in the cybersecurity domain. For example, the URL NLE type also occurs frequently in other domains, but might not have the similar information value of performing MLM as the cybersecurity domain. Therefore, the decision to do MLM on URL tokens could depend on the domain. An example of a domain where MLM of URL tokens might be inappropriate is in the Twitter text domain, where links are randomized by the Twitter URL shortener.

\section*{Acknowledgements}
\label{sec:ack}

This work was supported by Institute of Information \& Communications Technology Planning \& Evaluation (IITP)  grant funded by the Korea government (MSIT). (No. 2022-0-00740, The Development of Darkweb Hidden Service Identification and Real IP Trace Technology)

\bibliography{anthology,custom,references}
\bibliographystyle{acl_natbib}

\clearpage

\appendix

\section{Corpus}
\label{appendix:corpus}
We construct our corpus with data from the following sources:
\begin{itemize}
\item \textbf{Online Security Articles}: Following work in CTI~\cite{ZHAO2020101867} we identify news outlets, corporate blogs, and personal blogs that discuss cybersecurity content. We find a total of 60 online sources (see Appendix~\ref{sec:articles-sources}). We extract text using Scrapy\footnote{\url{https://scrapy.org}} and Selenium\footnote{\url{https://www.selenium.dev}}.

\item \textbf{Security Paper Abstracts}: We identify security conferences that make the abstracts of accepted papers publicly available. In total, we collect 7,301 abstracts from 8 different security conferences.

\item \textbf{Wikipedia Articles}: We start with the cybercrime category\footnote{\url{https://en.wikipedia.org/wiki/Category:Cybercrime}}, recursively visiting its sub-categories one by one and collecting all the pages under each category while discarding irrelevant pages through manual inspection. We collect a total of 3,411 pages.

\item \textbf{CVE Descriptions}: The CVE database\footnote{\url{https://cve.mitre.org/}} is a database of publicly disclosed vulnerabilities.
Each vulnerability is assigned an ID and given a short description.
We process the database to remove duplicate descriptions, incomplete (reserved or unused CVEs) descriptions, and descriptions that are too short (less than 10 words).
We retain a total of 184,956 CVE entries of unique and informative descriptions.

\end{itemize}

\section{Tokens for Probing Task}
\label{appendix:probing}
We collect a total of 556 phrases from the attack techniques and subtechniques listed in the MITRE database.
For simplification, we only select single-token target words.
From the phrases, we seperate into words and check if the word is in the RoBERTa tokenizer.
This way we find a total of 871 target tokens.
Since many common tokens such as `` and" or `` to" are selected in this way, we apply a simple filter by token IDs to remove common tokens (ID < 25,000).
This results in a target token list of 226 tokens including `` Unix" and `` runtime".

\section{Experiment settings}
\label{appendix:Experiment-settings}
\subsection{Pretraining}
To pretrain the models for Section~\ref{sec:pretrain}, we train on 2 NVIDIA A100 80GB GPUs. We use a slightly lowered effective batch size of 2024 to accomodate the smaller corpus size, and a warmup ratio of 0.048 following RoBERTa (which uses fixed steps). Other hyperparameters regarding including learning weights, weight decay, and adam hyperparameters are kept the same as RoBERTa's.

To pretrain the full \model{} model described in Section~\ref{sec:cybert}, we train on 4 NVIDIA A100 80GB GPUs on our full corpus. The hyperparameter settings are kept the same as above with the exception of the maximum epochs, which is set to 200.

\subsection{Fine-tuning}

For all tasks, fine-tuning is done with 20 max epochs, warmup ratio of 0.06, and an early stopping patience of 4 based on evaluation loss on the dev set. For the token classification tasks, evaluation is done every epoch. For sequence classification tasks, evaluation is done every 200 steps. For the multi-choice QA task, evaluation is done every 200 steps.

To best compare the models themselves, we keep implementations simple for downstream tasks.
We use the Hugging Face~\cite{wolf2020huggingfaces} implementations of each task.
For token classification, sequence classification, and multichoice QA tasks, we use the AutoModelForTokenClassification, AutoModelForSequenceClassification, and AutoModelForMultipleChoice, respectively.

To simplify hyperparameter selection, we select a batch size for each task following observations that input types varied heavily on the task (such as sentences and documents).
First we conducted a grid search of learning rates $\in$ \{2e−5, 3e−5, 5e−5, 1e-4\} and batch sizes $\in$ \{1, 2, 4, 8, 32\} for the CyNER task.
We identified learning rate of 5e-5 and batch size 32 worked best. This combination worked well with all tasks involving single-sentence inputs. CYDEC and TwitterThreats use this setting.
For CySecED (document), we did grid search of batch sizes $\in$ \{1, 2, 4, 8, 32\} and find that batch size of 1 works best. CASIE also uses this setting.
For MTDB (QA), we did grid search of batch sizes $\in$ \{1, 2, 4, 8, 32\} and find that batch size of 8 works best.

We use the train/eval/test given in the datasets if possible (CyNER, CySecED, MTDB). If the dataset does not have splits (CASIE, TwitterThreats, CyDEC), we split randomly at a 8:1:1 ratio.
Since CyNER deals with identifying exact spans while CySecED and CASIE deals with identifying event triggers, we use a stricter matching for CyNER (with seqeval\footnote{\url{https://github.com/chakki-works/seqeval}}) but use a loose matching scheme for CySecED and CASIE.

\section{Article Sources}
\label{sec:articles-sources}
In Table~\ref{tab:datacollection-datalist}, we list detailed online data sources from which we collect cybersecurity domain text.

\begin{table}[!t]
\renewcommand{\arraystretch}{1.1}
\resizebox{\columnwidth}{!}{
\begin{tabular}{llr}
\toprule
    \textbf{Source} & \textbf{Type} & \textbf{\# Pages Collected} \\
    \midrule
    InfoSecurity & News & 23,217 \\
    ThreatPost & News & 15,742 \\
    The Hacker News & News & 10,049 \\
    Bleeping Computer* & News & 8,852 \\
    Infosec Institute & News & 6,086 \\
    Security Intelligence & News & 1,824 \\
    The Record & News & 1,471 \\
    Cyber Security Hub & News & 902 \\
    \midrule
    Schneier on Security & Personal Blog & 8,008 \\
    TaoSecurity Blog* & Personal Blog & 3,044 \\
    Krebs on Security & Personal Blog & 2,151 \\
    Darknet & Personal Blog & 2,081 \\
    Ddanchev Blog* & Personal Blog & 1,575 \\
    hpHosts Blog* & Personal Blog & 1,057 \\
    Hexacorn Blog & Personal Blog & 784 \\
    Garwarner Blog* & Personal Blog & 570 \\
    Kahu Security* & Personal Blog & 194 \\
    SkullSecurity* & Personal Blog & 144 \\
    Carnal0wnage* & Personal Blog & 124 \\
    SecNiche* & Personal Blog & 94 \\
    DeepEnd Research* & Personal Blog & 23 \\
    \midrule
    Naked Security & Corporate Blog & 13,233 \\
    State of Security* & Corporate Blog & 5,233 \\
    WeLiveSecurity & Corporate Blog & 5,186 \\
    Palo Alto Networks & Corporate Blog & 3,482 \\
    Malwarebytes & Corporate Blog & 3,359 \\
    Securosis & Corporate Blog & 3,302 \\
    Microsoft & Corporate Blog & 2,902 \\
    Securelist & Corporate Blog & 2,897 \\
    Sophos* & Corporate Blog & 1,987 \\
    Sucuri* & Corporate Blog & 1,718 \\
    MSRC & Corporate Blog & 1,473 \\
    Spider Labs* & Corporate Blog & 1,463 \\
    Webroot* & Corporate Blog & 1,429 \\
    Recorded Future & Corporate Blog & 1,280 \\
    Zscaler* & Corporate Blog & 782 \\
    Unit42* & Corporate Blog & 771 \\
    NETSCOUT & Corporate Blog & 731 \\
    Radware & Corporate Blog & 720 \\
    Trustwave Blog & Corporate Blog & 676 \\    
    Forcepoint* & Corporate Blog & 665 \\
    SecureAuth & Corporate Blog & 583 \\
    Trend Micro (News)* & Corporate Blog & 494 \\
    Cloudflare & Corporate Blog & 449 \\
    Infoblox* & Corporate Blog & 403 \\
    BitDefender & Corporate Blog & 400 \\
    Honeynet Project* & Corporate Blog & 395 \\
    Mandiant* & Corporate Blog & 355 \\
    CoreSecurity & Corporate Blog & 257 \\
    Intezer* & Corporate Blog & 236 \\
    Symantec Enterprise Blogs* & Corporate Blog & 219 \\
    LookingGlass & Corporate Blog & 214 \\
    Veracode & Corporate Blog & 203 \\
    SEI (CERT/CC)* & Corporate Blog & 174 \\
    FireEye* & Corporate Blog & 148 \\
    CrowdStrike* & Corporate Blog & 144 \\
    Trend Micro (Research)* & Corporate Blog & 141 \\
    Juniper* & Corporate Blog & 122 \\
    Fox IT* & Corporate Blog & 109 \\
    Verisign Blog & Corporate Blog & 100 \\
    \bottomrule
\end{tabular}%
}
\caption{Full list of security news articles and security blogs used for corpus collection. The sources included in our pretraining subset are marked by *.}
\label{tab:datacollection-datalist}
\end{table}



\end{document}